\documentclass{article}
\usepackage{jheppub}
\usepackage[toc,page]{appendix}
\usepackage{graphicx}
\usepackage{subcaption}
\usepackage{amsmath,amssymb,amscd,amsfonts,mathtools}
\usepackage{slashed}

\def\ba{\begin{eqnarray}}
\def\ea{\end{eqnarray}}
\def\be{\begin{equation}}
\def\ee{\end{equation}}

\renewcommand{\Im}[1]{\text{Im}{#1}}
\renewcommand{\Re}[1]{\text{Re}{#1}}

\title{Holographic description of an anisotropic Dirac  semimetal}

\author[a]{Sebasti\'an Bahamondes}
\author[b]{Ignacio Salazar Landea}
\author[a]{Rodrigo Soto-Garrido}

\affiliation[a]{Facultad de F\'isica, Pontificia Universidad Cat\'olica de Chile, Vicu\~{n}a Mackenna 4860, Santiago, Chile}\affiliation[b]{Instituto de F\'\i sica de La Plata - CONICET, C.C. 67, 1900 La Plata, Argentina}

\emailAdd{sbahamondes@uc.cl}
\emailAdd{peznacho@gmail.com}
\emailAdd{rodrigo.sotog@gmail.com}

\abstract{Holographic quantum matter exploits the AdS/CFT correspondence to study systems in condensed matter physics. An example of these systems are strongly correlated semimetals, which feature a rich phase diagram structure. In this work, we present a holographic model for a Dirac semimetal in $2+1$ dimensions that features a topological phase transition. Our construction relies on deforming a relativistic UV fixed point with some relevant operators that explicitly break rotations and some internal symmetries. The phase diagram for different values of the relevant coupling constants is obtained. The different phases are characterized by distinct dispersion relations for probe fermionic modes in the AdS geometry. We find semi-metallic phases characterized by the presence of Dirac cones and an insulating phase featuring a mass gap with a mild anisotropy. Remarkably, we find as well an anisotropic semi-Dirac phase characterized by a massless a fermionic excitation dispersing linearly in one direction while quadratically in the other. 
}

\begin{document}
\maketitle

\section{Introduction}

The discovery of graphene sparked a great interest towards a better understanding of the behavior of relativistic fermions in $2+1$ dimensions. These fermions describe the low energy effective excitations around special points of the  Brillouin zone and were found/predicted for a number of systems. Sometimes, the position of Dirac cones in momentum space may be changed by a band tuning parameter. Fine tuning this parameter the Dirac cones might merge  \cite{montambaux2009merging} 
and this merging results into a topological phase transition between a semi-metallic Dirac state and a band insulator featuring a band gap. When these two phases meet we have two possibilities for the topological critical point: either it disperses quadratically in both directions or it disperses quadratically in one direction while linearly in the other, featuring an anisotropic Dirac semimetal \cite{banerjee2009tight}. For instance, this kind of behavior was observed experimentally in black phosphorus \cite{kim2015}, where the dispersion is linear along armchair direction and it is quadratic in the zigzag direction of the underlying honeycomb lattice. Other systems where this anisotropic dispersion was proposed include TiO$_2$/VO$_2$ nanostructures under confinement \cite{Pardo2009} and photonic metamaterials \cite{wu2014}

Understanding the quantum field theoretical description of these non-relativistic critical points with interactions using standard methods as the $\epsilon$ expansion or the large $N$ limit give ambiguous results depending on the way limits are taken \cite{isobe2016emergent,roy2018quantum,sur2019unifying,Uryszek:2019joy}. In this context one may wonder if the AdS/CFT duality might give some insightful understanding via an alternative approach.
Over the last decades, the holographic duality (also known as the AdS/CFT and gauge/gravity duality) has been used to study several condensed matter systems in the strong coupling regime. For a review of the last developments see  \cite{hartnoll2018,Zaanen-2015,Zaanen:2021}. In the context of topological phase transitions in $3+1$ dimensions we have gravitational holographic duals for Weyl semimetals \cite{Landsteiner:2015lsa,Landsteiner:2015pdh, Landsteiner:2016stv,Ji:2021aan}, multi-Weyl semimetals \cite{Dantas:2019rgp,Juricic:2020sgg} and nodal line semimetals \cite{Liu:2018bye} among others.
On the other hand, $2+1$ dimensional topological phase transitions featuring quadratic band touching were recently studied in \cite{Grandi:2021bsp,Grandi2022,Grandi:2023jna}.

In this paper we propose an holographic construction featuring a topological phase transition characterized by a Dirac semimetal at its critical point. To do so we will start by considering a free toy model in Section \ref{sect:free} featuring the desired anisotropic dispersion relation. We will study the symmetry breaking pattern of the free model, and mimic it for the boundary conditions of some classical fields living outside an $AdS$ black hole, in Section \ref{sect:model}, where we introduce our holographic model. Then, we will study the dynamics of probe fermions living on top of our holographic model in Section \ref{sec:free_fermions}. In Section \ref{sec:pole_skipping} we take a slight detour to probe our numerical algorithm against previously reported analytic results for pole-skipping frequencies. In this way, we confirm that the near horizon analysis presented in \cite{Ceplak:2020} is consistent with the full fermionic Green function. We then turn to the main results of this paper in Section \ref{sec:results}. These consist on the study of the fermionic Green function in our holographic model and specially on the identification of an anisotropic dispersion relation dispersing linearly in one direction while quadratically in the other. We conclude by enumerating some likely future directions in Section \ref{sect:conclusion}. Appendix \ref{sec:appendix_A} shows some details on boundary conditions and its relation to pole-skipping.

\section{A free fermionic model}
\label{sect:free}
Our starting point will be a free fermionic model featuring the topological phase transition we are interested in. From this model we will extract a symmetry breaking pattern associated to the construction that we will use as inspiration for our holographic  model. We should enphasize here that the holographic model will not be dual to this free fermionic model, but we are only using it to understand the underlying symmetries associated to our construction.

\begin{figure}[!htb]
    \centering
    \includegraphics[width=\textwidth]{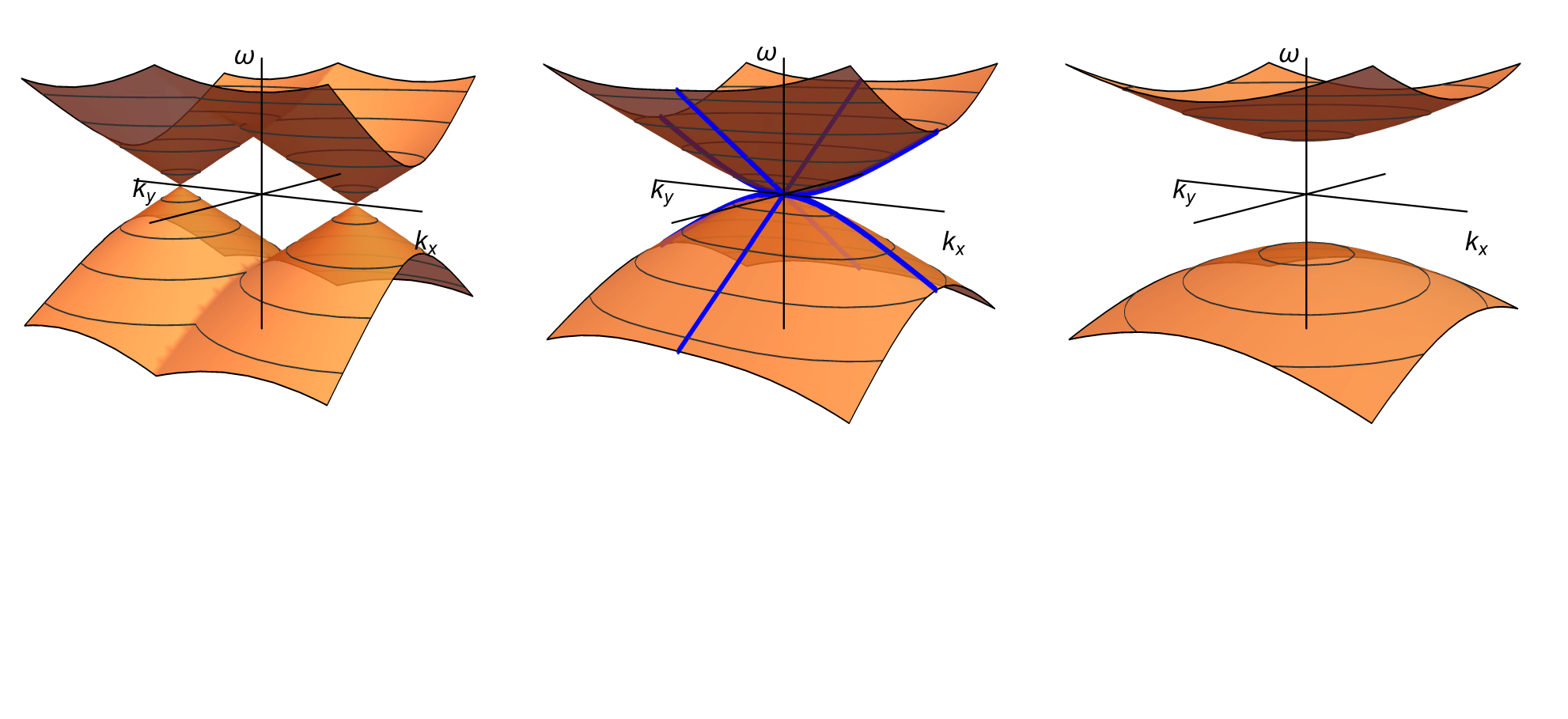}
    \vspace*{-3.1cm}
    \caption{Dispersion relations for low energy quasi-particle excitations (QPE's) for all different choices of coupling parameters $\Delta_1$ and $\Delta_2$. The left-most plot corresponds to the case in which $\Delta_1 > \Delta_2$, when this graphene-like system behaves like a semimetal. The middle plot shows the anisotropic semi-Dirac phase-transition ($\Delta_1 = \Delta_2$) in which the energy bands behave quadratically for $k_x$ and linearly for $k_y$. The right-most plot shows the remaining case, when $\Delta_1 < \Delta_2$, resulting in a gapped system.}
    \label{fig:pretty_dispersions}
\end{figure}

Let us start with a simple free fermionic model in $2+1$ dimensions that consists on two Dirac fermions in the UV coupled via the gausian relevant deformations $\Delta_1$ and $\Delta_2$ \cite{Pena-Benitez:2018dar}
\begin{equation}
    H_2= H_D \otimes \sigma_0+ \left( \Delta_1\sigma_1 \otimes \sigma_1+ \Delta_2\sigma_3 \otimes \sigma_3  \right),
\end{equation}
where we denote the Pauli matrices by $\sigma_{j}\quad (j=1,2,3)$, as well as $\sigma_0 = I_{2\times2}$.
As shown in Figure \ref{fig:pretty_dispersions}, whenever $\Delta_2>\Delta_1$ this system is gaped, while it features two Dirac cones separated in momentum when $\Delta_2<\Delta_1$.
The spectrum for $\Delta_1=\Delta_2\equiv\Delta$ reads
\begin{equation}\label{eq:anysotropic_relation}
    \omega=\pm \sqrt{k^2+2 \Delta^2\pm 2\sqrt{k_x^2 \Delta^2+\Delta^4}}
\end{equation}
The conduction and the valence bands crossing at zero energy  features an  anisotropic dispersion relation given by:
\begin{equation}\label{eq:anisotropy}
    \omega\approx \pm\sqrt{ k_y^2 + \frac{k_x^4}{4\Delta^2}}
\end{equation}

Let us now write the action that describes these two Dirac fermions $\Psi=(\psi,\xi)$,
\begin{equation}
    S=\int d^3x \left(i\bar \psi \gamma^a\partial_a \psi+i\bar \xi \gamma^a\partial_a \xi+\Delta_2 \left( \bar\psi\psi-\bar\xi\xi \right)-\Delta_1\left( i\bar\psi\gamma^1\xi+i\bar\xi\gamma^1\psi \right)  \right)
\end{equation}
where $\gamma^a=(\sigma_3,-i \sigma_2,i \sigma_1)$.

Interestingly this action is equivalent to coupling $\Psi$ to a constant $SU(2)$ gauge field $A_x^{(1)}=\Delta_1$ and a constant scalar field transforming in the adjoint representation of $SU(2)$, $\varphi=\Delta_2 \sigma_3$.

Next we proceed to construct an holographic model featuring these dispersion relations. Our procedure will be the standard bottom up approach. We identify the symmetries and the scheme of symmetry breaking of our toy free model. Then we gauge those symmetries in one extra dimension and also add a scalar doublet to realize the symmetry breaking pattern described above.
Hence a very minimal choice  will consist on studying the dynamics of a $SU(2)$ gauge field  and a scalar in the adjoint representation of $SU(2)$ living in a fixed AdS black hole background.

\section{Holographic model}
\label{sect:model}

Inspired in our free model in the previous section, we will propose that a minimal construction in the probe limit will consist on a scalar fields transforming in the adjoint representation of $SU(2)$ coupled to a gauge field, both living on top of a fixed AdS black hole metric
\begin{gather}
S_{\mathrm{b}}= \int\!\mathrm{d}^4x\,\sqrt{-g}\left[\mathrm{Tr}\left(\left(D^\mu\Phi\right)^\dagger\left(D_\mu\Phi\right)\right)+m^2\mathrm{Tr}\left(\Phi^\dagger\Phi\right)+
\frac{1}{4}\mathrm{Tr}\left(G_{\mu\nu}G^{\mu\nu}\right)\right],\label{eq:background_action}
\end{gather}
where $G$ is the strength of the $SU(2)$ gauge field defined from $B=B^j\,\sigma_j$ with $j=1,2,3$ in the standard way: $G=d{B} + i(q_e/2) {B}\wedge B$.  Also, since the scalar field $\Phi$ will be used in the adjoint representation of $SU(2)$, it corresponds to a $2\times 2$ matrix, and therefore the covariant derivative acts in the corresponding representation as:
\begin{eqnarray}\label{eq:adjoint_cov_dev}
D_\mu\Phi=\nabla_\mu\Phi +i q_e\left[\sigma_a B_a,\Phi\right] + iq_e A_\mu\Phi,
\end{eqnarray}
where one would consider naively that $q_e=2q_f=2q$ taking for instance the intuition from Landau's theory of superconductivity, although that might not be the case here. A similar model, with different boundary conditions, was studied in \cite{Giordano:2015vsa}.

As stated before, we work in the probe-limit, where the background fields $\Phi$ and $B$ do not backreact on the space-time metric. Therefore, the black-brane $\mathrm{AdS}_4$-Schwarzschild solution can be imposed for the Einstein equations 
\begin{equation}\label{eq:metric}
    \mathrm d s^2 = \frac{1}{r^2}\left(-f(r)\mathrm dt^2 + dx^2 + dy^2+ \frac{dr^2}{f(r)} \right)
\end{equation}
where Poincaré coordinates have been used (the AdS boundary is at $r=0$), and the $\mathrm{AdS}$ radius $L$ has been set to $1$. Also, setting the black-brane event horizon at $r=1$, the blackening factor is $f(r) = 1-r^3$ and the temperature reads $T = \frac{3}{4\pi}$.

The equations of motion (EOM's) for the scalar and gauge fields are the covariant Klein-Gordon and Yang-Mills equations given by
\begin{align}
    (D_\mu D^\mu-m^2)\Phi &= 0 \nonumber\\
    D_\mu G^{\mu\nu} &= iq\left(\left[\Phi^\dagger,D^\nu\Phi\right]-\left[\Phi^\dagger,D^\nu\Phi\right]^\dagger\right).\label{eq:yangmills}
\end{align}

Now we will turn on a relevant deformation that breaks the boundary $SU(2)$ group in the precise way described in Section \ref{sect:free} and we let the corresponding fields to evolve into the bulk while remaining homogeneous in the boundary space-time directions. This translates into the following \textit{ansatz} and boundary conditions
\begin{gather}
    \Phi = \phi(r)\sigma_3 \xrightarrow[r\to 0]{} r\Delta_2\sigma_3 \,,\nonumber \\
    B = B(r)\sigma_1dx\xrightarrow[r\to0]{}\Delta_1\sigma_1dx\,.
    \label{eq:gauge_ansatz}
\end{gather}

Using the ansatz given in  ~\eqref{eq:gauge_ansatz} and the metric in \eqref{eq:metric}, the equations of motion obtained from  \eqref{eq:yangmills} read
\begin{align}
(4r^2B(r)^2-2)\phi(z)-r\left[r\frac{\mathrm{d}f}{\mathrm{d}r}\frac{\mathrm{d}\phi}{\mathrm{d}r}+f(r)\left(r\frac{\mathrm{d}^2\phi}{\mathrm{d}r^2}-2\frac{\mathrm{d}\phi}{\mathrm{d}r}\right)\right]&=0\nonumber\\
8B(r)\phi(r)^2-r^2\left(\frac{\mathrm{d}B}{\mathrm{d}r}\frac{\mathrm{d}f}{\mathrm{d}r}+f(r)\frac{\mathrm{d}^2B}{\mathrm{d}r^2}\right)&=0\label{eq:gauge_eom}
\end{align}

Notice that the mass of the scalar field has been set to $m^2 = -2$, so that the dual boundary scalar operator has dimension $2$. This is done so the aforementioned dual operator has the same dimension as a fermion mass term operator $\overline{\psi}\psi$, the would be naive scaling dimension we expect for these couplings.

\begin{figure}[!htb]
    \centering
    \includegraphics[width=1\textwidth]{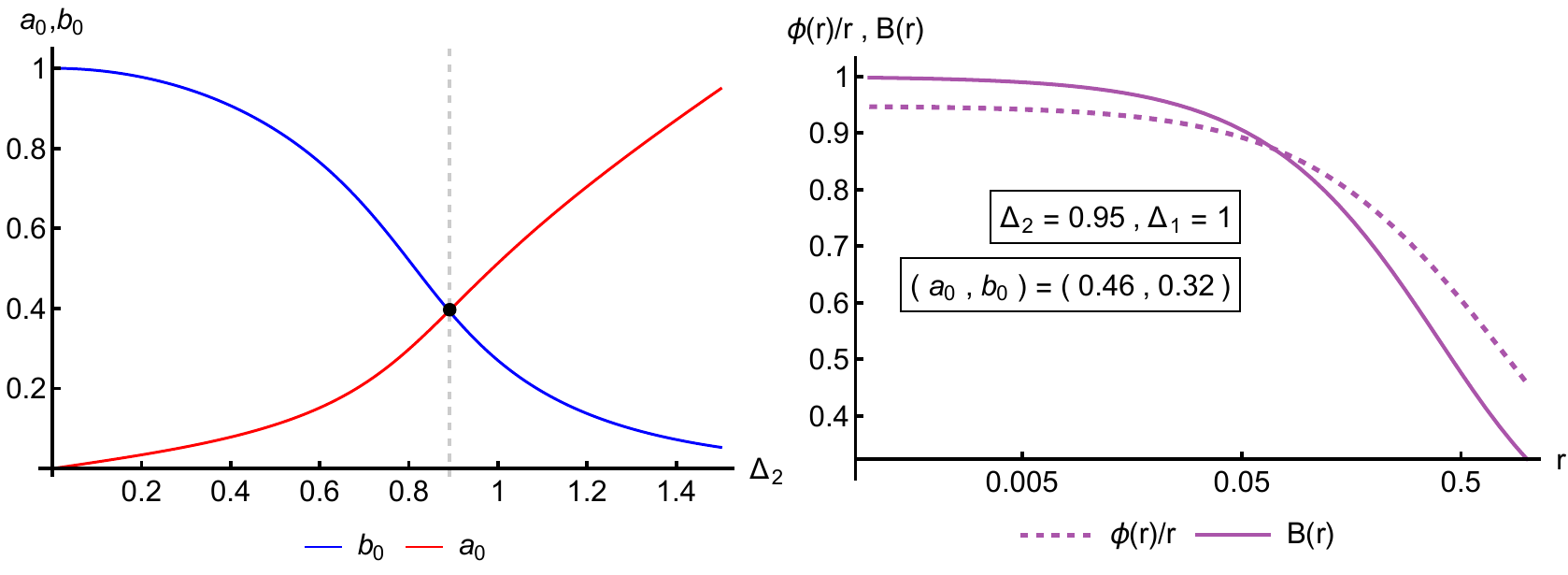}
    \caption{Left plot: Dependence of the shooting parameters $a_0$ and $b_0$ on $\Delta_2$, for fixed $\Delta_1 = 1$. The existence of a critical value for $\Delta_2$ is insinuated by the point of crossing of $a_0$ and $b_0$ at $\Delta_2 =\Delta_{2}^{(c)}\approx 0.8831\ldots$ Right plot: Profiles for the background bosonic fields at a sample value for the shooting parameters $a_0$ and $b_0$ (and their corresponding values for the UV boundary conditions for the background fields, $\Delta_1$ and $\Delta_2$). The $r$-axis is set in logarithmic scale. Notice that the choice to plot $\phi(r)/r$ is made so as to explicitly see the value of $\Delta_2$ as the intersection of the dashed curve with the vertical axis.}
    \label{fig:boson_fields}
\end{figure}

We now proceed to integrate numerically equations  \eqref{eq:gauge_eom}  using a shooting method from a regular near horizon
\begin{align}
    \phi_\text{IR}(r) &= a_0+a_1(1-r)+a_2(1-r)^2+a_3(1-r)^3+\cdots\nonumber\,,\\
    B_\text{IR}(r) &= b_0+b_1(1-r)+b_2(1-r)^2+b_3(1-r)^3+\cdots\label{eq:IR_gauge_ansatz}\,,
\end{align}
towards  UV boundary conditions given by ~\eqref{eq:gauge_ansatz}. A typical profile is shown in Figure \ref{fig:boson_fields}.
We will interpret the values of the fields at fixed radial coordinate $r$ as the renormalized values of the deformation parameters $\Delta_i$ as we run along the RG flow. Hence we expect that the topological phase transition will occur at the point $\phi(r_h)=B(r_h)$, following the intuition from \cite{Landsteiner:2015lsa}. We will see this is indeed the case when we probe our black holes with some bulk fermions.%

\section{Fermions in the AdS bulk}\label{sec:free_fermions}

In order to test our minimal model we will add some fermionic probe fields to the bulk coupled to our metric and matter field, to check if the dual dispersion relations follow the expected behaviors in the different regimes.

To do so we will consider a fermion $SU(2)$ doublet $\Psi = \left(\begin{array}{c}
     \psi_1  \\
     \psi_2 
\end{array}\right)$. In this framework, it corresponds to an 8-tuple, since the gauge interaction introduces an additional index to the 4-dimensional Dirac spinors $\psi_i$ \cite{Giordano2017,Grandi2022}. We will minimally couple this Dirac fermion to our $SU(2)$ gauge field and we will also incorporate a Yukawa coupling to the adjoint scalar

\begin{equation}\label{eq:full_action}
S_f=i\int\!\mathrm{d}^4x\left(\overline{\Psi}\slashed{D}\Psi+\lambda\overline{\Psi}\Phi\Psi\right)\,.
\end{equation}
Here the spinorial covariant derivative has both a gravitational and a gauge sector
\begin{equation}\label{eq:spin_cov_dev}
    D_\mu=\left(I_{4\times 4}\otimes I_{2\times 2}\right)\nabla_\mu+\left(\Gamma_\mu\otimes I_{2\times 2}\right)+\left(I_{4\times 4}\otimes iq_fB_\mu^{a}\sigma_a\right),
\end{equation}
where $\Gamma_\mu$ are the affine connections for this geometry, which couple the spinors to the curved gravitational background. Notice that a different coupling charge $q_f$ has been implemented for fermions. 

Given the background geometry of the bulk in Equation \eqref{eq:metric}, the appropriate set of \textit{vielbeins} is given by:
\begin{align*}
    e_0=-r\frac{\partial_t}{\sqrt{f(r)}}\;,\;
    e_1=r \partial_x\;,\;e_2=r \partial_y\;,\;e_3=r\sqrt{f(r)} \partial_r.
\end{align*}
To write down the Dirac equation, one must go through the standard procedure of calculating the spin connections $\omega_{ab\mu}$ from these \textit{vielbeins}, to obtain the $\Gamma_\mu$ matrices through $\Gamma_\mu = \frac{1}{8}\omega_{ab\mu}\left[\gamma^a,\gamma^{b}\right]$, where $\gamma^a$ are a certain representation of the Clifford algebra in 4 dimensions. 

The Dirac equation for the massless fermions reads
\begin{equation}
    \left(\slashed{D}+\lambda\Phi\right)\Psi=0,
\end{equation} 
Notice that the representation of the Dirac $\gamma$-matrices with Greek indices corresponds to the curved version of the flat matrices taken in the following representation 

\begin{gather}
    \gamma^0 = \begin{bmatrix}
        0 & i\sigma_2\\
        i\sigma_2&0
    \end{bmatrix} \quad , \quad \gamma^1 = \begin{bmatrix}
        0 & \sigma_1\\
        \sigma_1 & 0
    \end{bmatrix}
    \quad , \quad \gamma^2 = \begin{bmatrix}
        0 & \sigma_3\\
        \sigma_3&0
    \end{bmatrix}
    \quad , \quad 
    \gamma^3 = \begin{bmatrix}
        -I_{2\times 2} & 0\\
        0 & I_{2\times 2}.
    \end{bmatrix}
\end{gather}

 From this last equation, we must obtain the boundary values of the fermion fields to calculate the retarded fermionic correlator. To holographically encode fermions, we must go into momentum space by writing $\Psi(t,r,x,y)=\left[\begin{array}{c}
     \Psi_1(r)\\
     \Psi_2(r)
\end{array}\right]e^{i(\omega t-k_xx-k_yy)}$, and then project the $8$-tuple $\Psi$ onto the eigen-space of the radial $\gamma^3$ matrix \cite{Gubser2010,Giordano2017}. Naming the two resulting 4-spinors $\Psi_{\pm}$ respectively, we finally perform the re-scaling $\Psi_{\pm}:=r^{3/2}f(r)^{-1/4}\zeta_{\pm}$ \cite{Grandi2022}. Taking the \textit{ansatz} for the background fields \eqref{eq:gauge_ansatz} into account  we obtain two coupled first-order ODE's
\begin{align}
    \frac{\mathrm{d}\zeta_+}{\mathrm{d}r}+\frac{i}{\sqrt{f(r)}}U\zeta_-=-\frac{\lambda}{r}\frac{\phi(r)}{\sqrt{f(r)}}\gamma^3\zeta_+ \label{eq:Dirac_eq_1}\\
    \frac{\mathrm{d}\zeta_-}{\mathrm{d}r}-\frac{i}{\sqrt{f(r)}}U\zeta_+=\frac{\lambda}{r}\frac{\phi(r)}{\sqrt{f(r)}}\gamma^3\zeta_-\label{eq:Dirac_eq_2}, 
\end{align}
where the operator $U$ has been defined as:
\begin{equation}\label{eq:operator_u}
    U(r;\omega,k_x,k_y)=\left[\begin{array}{cccc}
        k_y & k_x-\frac{\omega}{\sqrt{f(r)}} & 0 & q_fB(r) \\
        k_x + \frac{\omega}{\sqrt{f(r)}} & -k_y & q_fB(r) & 0 \\
        0 & q_fB(r) & k_y & k_x-\frac{\omega}{\sqrt{f(r)}} \\
        q_fB(r) & 0 & k_x+\frac{\omega}{\sqrt{f(r)}} & -k_y 
    \end{array}\right]
\end{equation}

 The Dirac equation must be solved, of course, with appropriate boundary conditions in the IR and the UV. First, the conditions in the IR are the standard in-falling boundary conditions at the black hole horizon $r = 1$ \cite{Grandi2022,Iqbal2009,hartnoll-2009}:
\begin{equation}\label{eq:infalling_bcs}
    \zeta_{\pm}(r) = F_\pm(r;\omega,k_x,k_y)(1-r)^{-i\omega/3},
\end{equation}
where the function $F_\pm(r;\omega,k_x,k_y)$ is an undetermined function that must be numerically solved by power series (up to a sufficiently high order) using the Dirac equation expansion near the horizon ($r = 1$). Since the Dirac equation is not analytic at $r=1$ due to the $\sqrt{f(r)}$ factors coming from the \textit{vielbeins}, this series solution will not be the typical Fröbenius-like infinite power series, but rather a half-integer power series expansion that requires much more terms for each order of approximation of $F_\pm$ \cite{Arnold:2013,Ceplak:2020}:

\begin{equation}\label{eq:half_int_expansion}
    \zeta_\pm^{\mathrm{IR}}(r) = (1-r)^{-i\omega/4\pi T}\left(a_0^{\pm}+a_{1/2}^{\pm}\sqrt{1-r}+a_{1}(1-r)+a_{3/2}^{\pm}(\sqrt{1-r})^3+a_2^{\pm}(1-r)^2+\cdots\right).
\end{equation}

The zeroth order imposing of these boundary conditions reads
\begin{equation}\label{eq:zeroth_order_IR}
    \zeta_{-,0}^{\mathrm{IR}}=-i(I_{2\times 2}\otimes\sigma_2)\zeta_{+,0}^{\mathrm{IR}}.
\end{equation}
 Therefore, the present system only has $4$ undetermined constants that can be chosen arbitrarily \textit{a priori}.

In the UV we have, on the other hand:
\begin{equation}\label{eq:retarded_correlator}
    \zeta_\pm^{\mathrm{UV}}(r) = \zeta_{\pm,0}^\mathrm{UV}+\mathcal{O}(r).
\end{equation}
According to the $\mathrm{AdS}/\mathrm{CFT}$ dictionary, the leading term of either of the fermion fields $\zeta_{\pm}$ in the UV can be chosen as the source of the dual fermionic operator, while the leading term of the other will be interpreted as the response \cite{Iqbal2009,Iqbal2011}.  The dispersion relation of fermion quasi-normal modes $\omega\equiv \omega(k_x,k_y;\Delta_1,\Delta_2)$ is obtained numerically from the poles of a matrix correlator $S(\omega,k_x,k_y;\Delta_1,\Delta_2)$ that relates the source and response of the fermion operator as

\begin{equation}\label{eq:fermion_correlator}
    \zeta_{+,0}^\mathrm{UV}=S(\omega,k_x,k_y;\Delta_1,\Delta_2)\zeta_{-,0}^\mathrm{UV}
\end{equation}
Since the Dirac equation is lineal, a linear relation is expected to hold between the fields in the deep IR and in the UV, through an undetermined matrix that depends on the parameters $\Delta_{1,2}$ and the momentum components, and that must be solved numerically \cite{Grandi2022}. Since the fields in the deep IR are related to each other through \eqref{eq:zeroth_order_IR}, a relation of the form $\zeta_{\pm}^{\mathrm{UV}} = M_{\pm}\zeta_{-,0}^\mathrm{
IR}$ can be numerically calculated, where $M_\pm$ are matrices that propagate the IR fields towards the UV, and that once again depend on the background and the fermions' momentum \cite{Grandi2022}. Using this fact, the correlator $S$ in eq.~\eqref{eq:fermion_correlator} can be expressed as 
\begin{eqnarray}
S=M_+M_-^{-1}    
\end{eqnarray}
and therefore the poles of the correlator can be read-off from the zeroes of the determinant of $M_-$ \cite{Grandi2022}. 
This numerical method has been exploited previously in the context of quadratic band-touching \cite{Grandi2022} and for non-abelian holographic superconductors \cite{Gubser2010}, so we apply it here to this new type of system to explicitly obtain the band structure of the boundary fermions, and its behavior through the phase transition that we expect to find.

 The correlator $S$ is related to the retarded fermionic Green's function in linear-response theory through the following identity \cite{Iqbal2009,Amon2010}
\begin{equation}
    S(\omega,k_x,k_y;\Delta_1,\Delta_2) = \gamma^0G_R(\omega,k_x,k_y;\Delta_1,\Delta_2),
\end{equation}
which allows for the numerical calculation of the full Green's function in the boundary once the numerical matrices $M_\pm$ have been obtained.

Finally, notice that after appropriate rescaling of eqs.\eqref{eq:Dirac_eq_1} and \eqref{eq:Dirac_eq_2}, the only free parameters remaining in the whole system are $q_f$ and $\lambda$. Both of these parameters determine the coupling strength of the fermions to the bosonic fields in the background. Once again, taking inspiration in Landau's theory we set $\lambda =q_f = 1/2$ for all the following numerical calculations. 

\section{Warm up: Fermionic pole skipping in the Schwarchild geometry }\label{sec:pole_skipping}

Pole skipping refers to a peculiar behavior of a Green function in momentum space where a line of poles meets a line of zeros \cite{Grozdanov:2017ajz,Blake:2017ris,Blake:2018leo}. In recent years it has caught a lot of attention as in certain situations the analytic structure of Green functions might be related to the chaotic behavior of the quantum system. At this stage we will detour for a moment from our main goal to test our numerical algorithm against recent analytical computations for the pole skipping points of the retarded fermionic Green's function on a Schwarchild AdS black hole \cite{Ceplak:2020}\footnote{This phenomenon is certainly not restricted to fermions in holographic black-hole backgrounds, and can be extended to more complicated settings (see for example \cite{Ahn:2024aiw,Baishya:2023xbj}).}. This is a particular case of the semimetal that is modeled by the action \eqref{eq:full_action} achieved by simply turning off the gauge and scalar fields, and only retaining the fermionic and gravitational sectors of the action. Hence it is the perfect warm up for our program and a numerical check of the results presented in \cite{Ceplak:2020} via a near horizon analysis. A near horizon analysis in our coordinate system is presented in Appendix \ref{sec:appendix_A}. 

\begin{figure}[!htb]
    \centering
    \includegraphics[width=.65\textwidth]{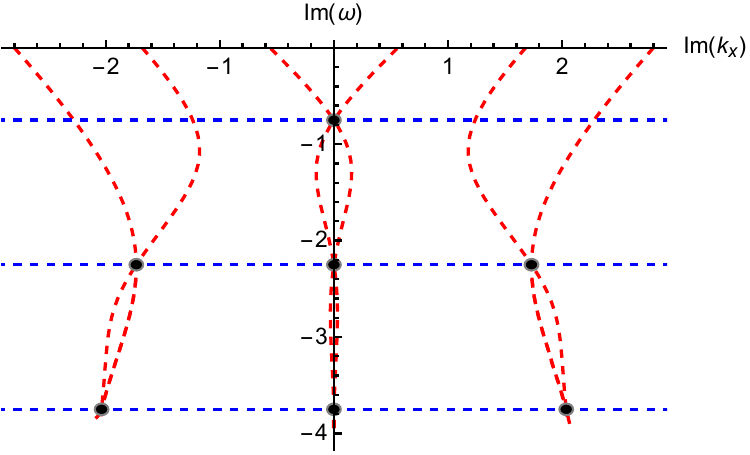}
    \caption{Plot of lines of poles (red lines) and lines of zeroes (blue lines) of the boundary Green's function in the $\mathrm{Im}(k_x)$-$\mathrm{Im}(\omega)$ plane, for the $\mathrm{AdS}_4$ Schwarzschild black hole background. As can be seen the pole-skipping points (black dots) are located precisely on the intersections of these two types of lines.}
    \label{fig:pole_skipping_plots}
\end{figure}

The first pole-skipping points read
\begin{align}\label{eq:first_pole_skips}
    \frac{\omega_1}{T} &= -\frac{3i}{4}\;,\; \frac{k}{T} = 0\nonumber \\
    \frac{\omega_2}{T} &= -\frac{9i}{4}\;,\; \frac{k}{T} = 0\, , \,\pm i\sqrt{3} \\
    \frac{\omega_3}{T} &= -\frac{15i}{4}\;,\; \frac{k}{T} = 0\, , \,\pm i\sqrt{\frac{3}{2}\left(5+\sqrt{5}\right)}\, , \,\pm i\sqrt{\frac{3}{2}\left(5-\sqrt{5}\right)}\nonumber\\
    &\qquad\vdots\nonumber
\end{align}
where frequencies shown in eq.~\eqref{eq:first_pole_skips} are just the first fermionic Matsubara frequencies and the associated pole-skipping spatial momenta $k$ are determined from the near-horizon physics of the Dirac equations \eqref{eq:Dirac_eq_1} and \eqref{eq:Dirac_eq_2}.  

The pole-skipping points are associated to values of the parameters $(\omega,k)$ for which the in-falling boundary conditions \eqref{eq:half_int_expansion} fail to give a unique solution to the Dirac equation, and instead result in two sets of free parameters, unrelated to one another, that makes the construction of the boundary Green's function ill-defined. The precise values shown in eq.~\eqref{eq:first_pole_skips} can be obtained by solving each coefficient in the expansion \eqref{eq:half_int_expansion} order by order, and looking for values of $(\omega,k)$ that make the resulting system of equations for each coefficient singular (see Appendix \ref{sec:appendix_A} for more details). The IR near-horizon physics that determines the pole-skipping points also translates into the analytic structure of the retarded correlator in the UV \eqref{eq:retarded_correlator}.

Specifically, the pole-skipping points $(\omega,k)$ correspond to intersections between lines of zeroes and lines of poles of the correlator in the purely imaginary $\omega-k$ plane \cite{Ceplak:2020}. 
Hence, by systematically setting $\Delta_1 = \Delta_2 = 0$, we can recover the pole skipping points by chasing the zeros of $\mathrm{det}(M_-)$ and $\mathrm{det}(M_+)$ defined in Section \ref{sec:free_fermions} that would correspond respectively to the poles and zeros of the retarded Green's function. The corresponding numerical results are shown in Figure \ref{fig:pole_skipping_plots}.

\section{Poles of the retarded Green's function}\label{sec:results}

\begin{figure}[!htb]
    \centering
    \includegraphics[width=0.65\textwidth]{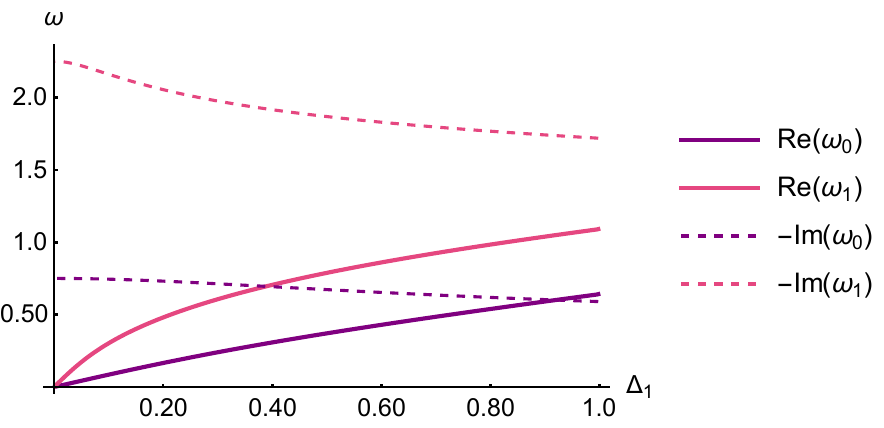}
    \caption{
    Evolution of the real and imaginary parts of the quasinormal frequencies of the lowest-lying and first excited modes, as a function of $\Delta_1$ for $\Delta_2 = 0$. As the fields are progresively turned off, both modes become massless, but their imaginary frequencies don't converge to the same value.}
\label{fig:zero_momentum_phase_transition}
\end{figure}

We now proceed to study the quasinormal modes associated to the fermionic retarded Green's function. Let us begin by restating the main features of our holographic model: we have a family of black holes depending on two independent parameters $\Delta_1$, $\Delta_2$ measured in units of the black hole temperature. On top of those black holes we set probe fermions to compute their dual fermionic Green's function. There will be two protagonist modes in our analysis: the lowest-lying mode of the boundary correlator, and the first excited mode. 
The frequencies of these two modes will be respectively denoted by $\omega_0$ and $\omega_1$ throughout the rest of this section. To better understand them let us first consider zero spatial momentum $k_x=k_y=0$ solutions. Turning off both sources $\Delta_1=\Delta_2=0$, we have a series of fermionic modes at $k_{x,y}=0$ with $\Re(\omega)=0$ and $\Im(\omega)\neq 0$. Now let us study what happens with these two modes when we turn on the $\Delta_1$ while keeping $\Delta_2=0$.  As we show in \ref{fig:zero_momentum_phase_transition} both modes acquire a real gap.

\begin{figure}[!htb]
    \centering
    \includegraphics[width=\textwidth]{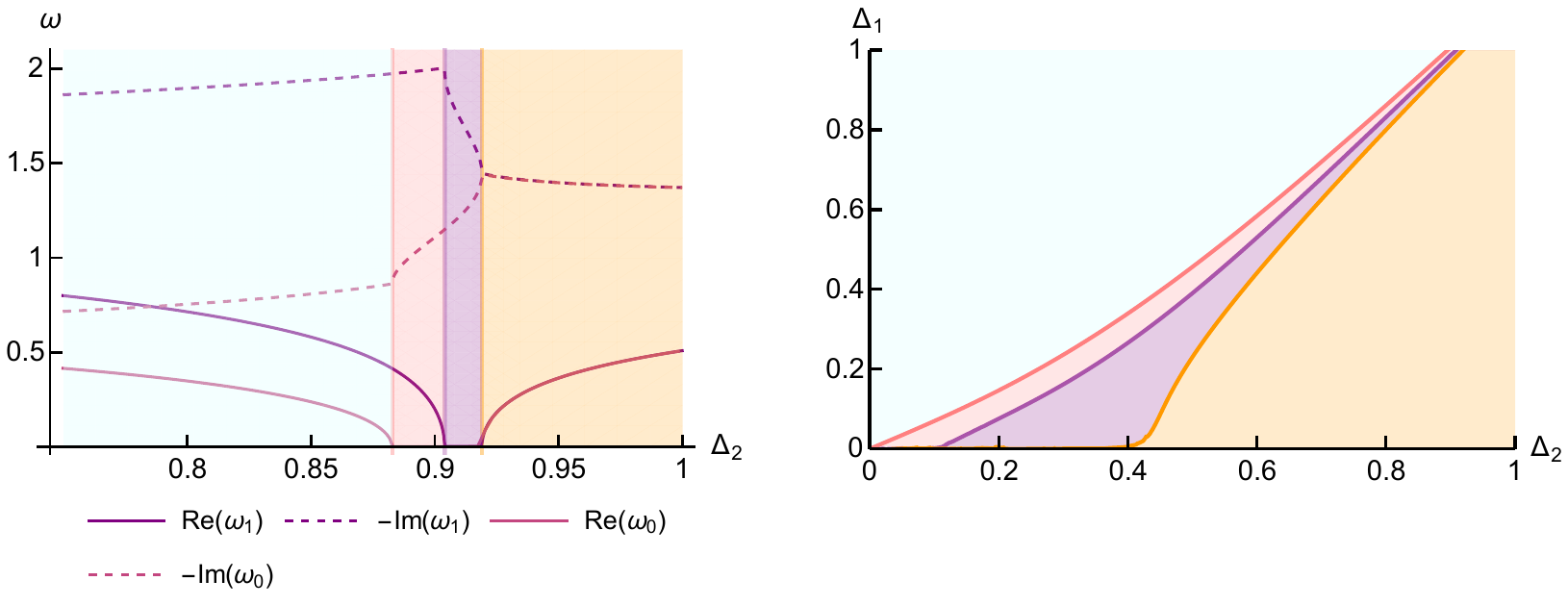}
    \caption{Left plot: Evolution of real and imaginary parts of the quasi-normal frequency at zero $k_{x,y}$ as a function of $\Delta_2$, for fixed $\Delta_1 = 1$. The frequency of the first excited mode is denoted by $\omega_1$, while that of the ground (lowest-lying) mode is denoted as $\omega_0$. Right plot: Phase diagram for the $\Delta_1-\Delta_2$ parameter space, where the different curves correspond to different phase transitions of the lowest and excited modes of the fermionic correlator, showing the different regions in the $\Delta_i$ plane defined by the transitions of the quasinormal frequencies for both $\omega_0$ and $\omega_1$, as shown in the left plot.} 
    \label{fig:phase_diagram}
\end{figure}

Now we are ready to turn on $\Delta_2$ while fixing $\Delta_1=1$. We see in Figure \ref{fig:phase_diagram} that the gap decreases for both modes as $\Delta_2$ increases, and eventually the lowest lying mode $\omega_0$ hits the real axes and becomes massless at $\Delta_2\approx 0.8831$. Notably, $\omega_0$ becomes null at $k_{x,y}=0$ exactly when the horizon values of the background fields $a_0$ and $b_0$ coincide  in Figure \ref{fig:boson_fields}. This fact  reinforces the interpretation of these being the corresponding renormalized IR values for the UV deformations $\Delta_i$. From our free model, our naive expectation would be that this mode would bounce again immediately. Contrary to that intuition we see that the mode remains massless for a finite range of $\Delta_2$. Eventually the first excited mode $\omega_1$ hits the real axis in the $\omega$ plane and we have a phase with two massless excitations. Finally, as $\Delta_2$ keeps increasing, $\omega_0$ and $\omega_1$ collide in the complex plane and, as a result, they merge and acquire a real mass gap.

\begin{figure}[!htb]
    \centering
    \begin{subfigure}{\textwidth}
    \centering
    \includegraphics[width=\textwidth]{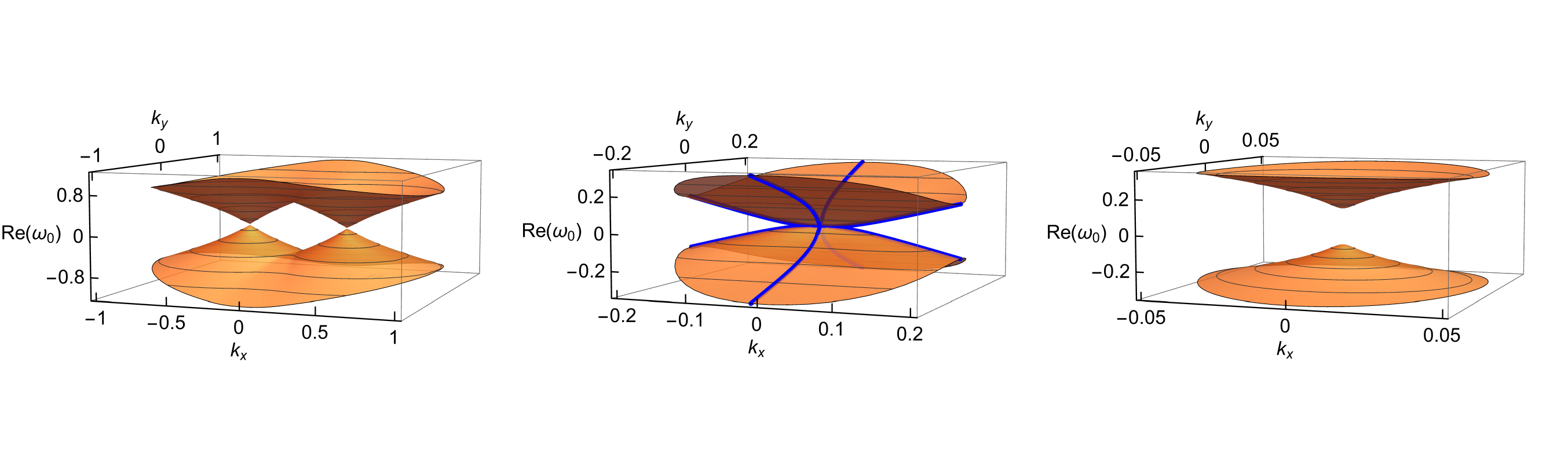}
    \vspace*{-1.8cm}
    \end{subfigure}
    \begin{subfigure}{\textwidth}
    \centering
        \includegraphics[width=\textwidth]{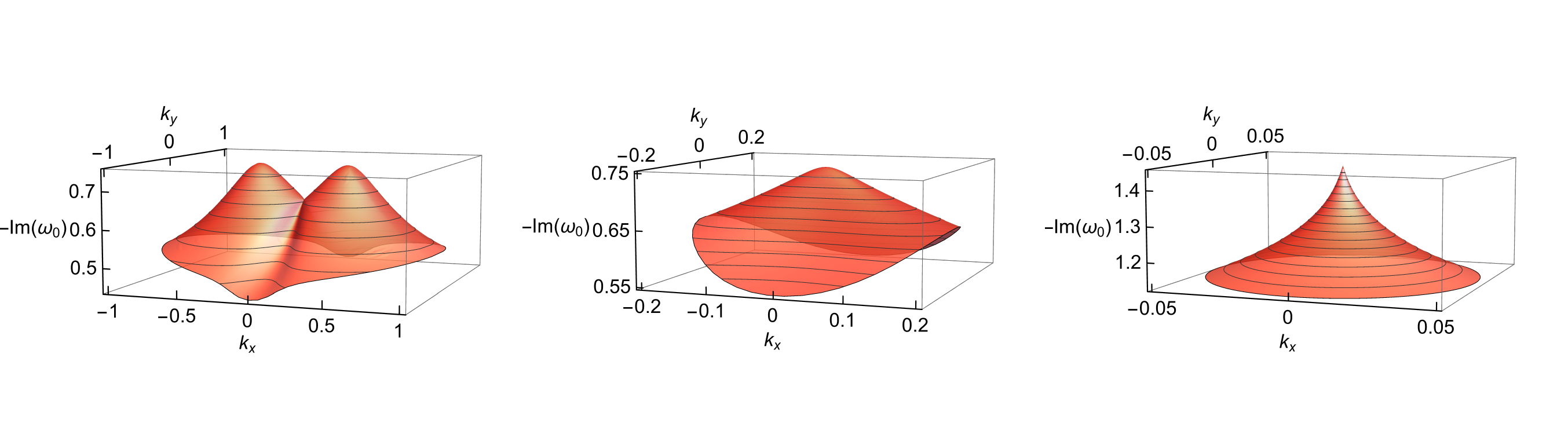}
    \end{subfigure}
    \caption{Plots of the full dispersion relation for the lowest lying mode $\omega_0$, for $\Delta_1 = 1$ and  $\Delta_2 \approx 0.6253,\, 0.8831,\,0.916 $ respectively. The plots in the top panel corresponds to the real part of $\omega_0$ for finite $k_{x,y}$, while plots in the bottom panel are the corresponding imaginary parts.}
    \label{fig:3D_plots}
\end{figure}

The  corresponding phase diagram in the $\Delta_1$-$\Delta_2$ plane with the different topological phases characterized by the corresponding fermionic excitations is shown in the right panel of Figure \ref{fig:phase_diagram}. To fully address the meaning of each of this phases we must turn on the momentum of the fermionic background fields. To this end we illustrate three particular dispersion relations in Figure \ref{fig:3D_plots}. 

We see that the real gap has a very different interpretation depending on which $\Delta_i$ dominates. 
For large $\Delta_2$ (orange region) we observe that we have a regular mass gap with a mild anisotropy. From \cite{Plantz:2018tqf}, we expect that the Yukawa coupling will gap these modes and this is indeed what happens. On the other hand, for small $\Delta_2$ (the light-blue region) the dispersion relation for the ground mode shows two Dirac cones separated one from the other in momentum space. This behavior could be expected as the gauge field contribution dominates in this region (see for instance \cite{Grandi:2021bsp}).

The red line located between the light-blue and pink regions corresponds to the transition line where both Dirac cones in the semi-metallic phase collide, and $\omega_0$ becomes null for $k_{x,y}=0$. Precisely at this point we see an anisotropic Dirac behavior, with the $x$ direction dispersing quadratically
 \begin{equation}
     \omega_{0}= -i\alpha+(\pm\beta-i\gamma)k_x^2 +\dots
 \end{equation}
while the $y$ direction disperses linearly
 \begin{equation}
     \omega_{0}= -i\alpha\pm v_f k_y+\dots,
 \end{equation}
where $v_f,\,\alpha,\,\beta,\,\gamma$ are parameters that are fitted to the numerical data. 

\begin{figure}[!htb]
    \centering
    \begin{subfigure}{\textwidth}
    \centering
 \includegraphics[width=.8\textwidth]{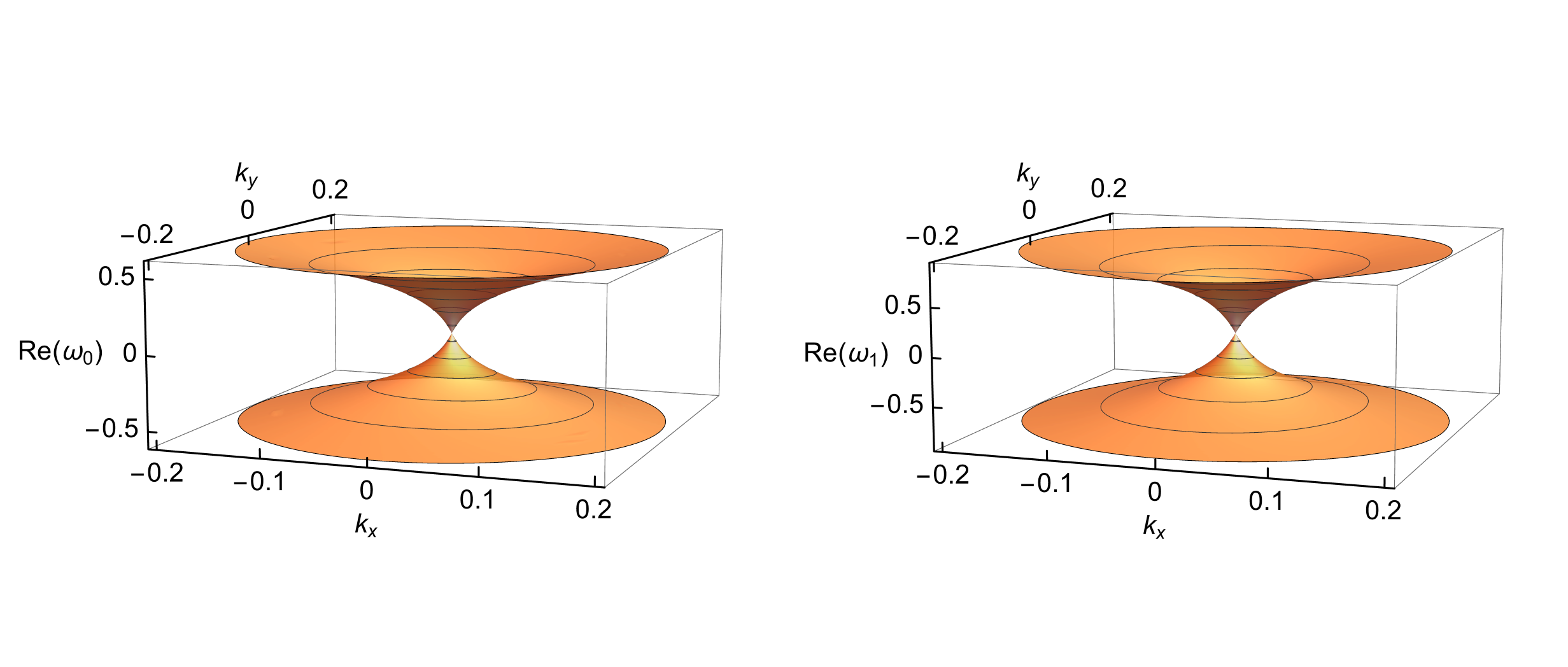}
    \vspace*{-1.6cm}
   \end{subfigure}
   \begin{subfigure}{\textwidth}
   \centering
       \includegraphics[width=.8\textwidth]{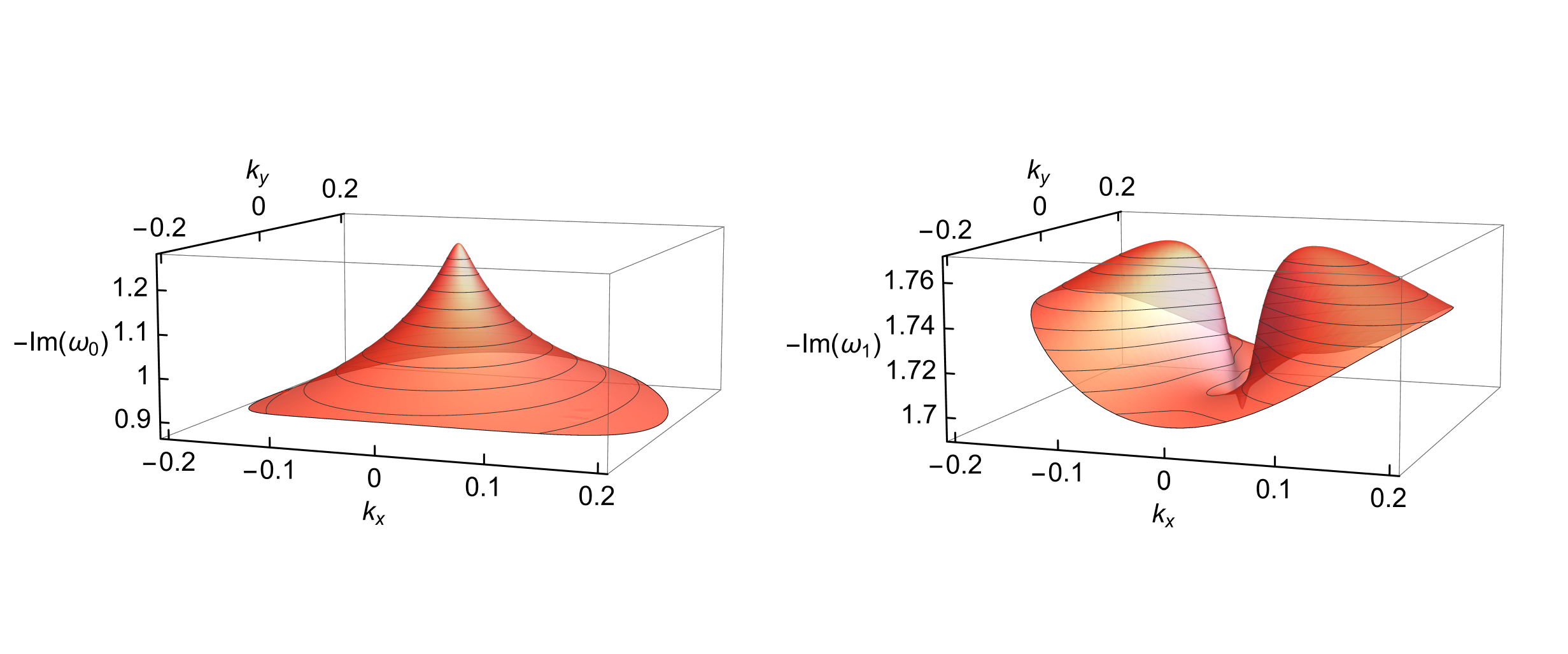}
   \end{subfigure}
     \caption{Top panel: Dispersion relation for the real part of $\omega_0$ (left plot) and $\omega_1$ (right plot) at $\Delta_2 \approx 0.9092$. Both modes showcase linear dispersion relations at small momenta, with the quadratic contribution to the $k_x$ direction of $\mathrm{Re}(\omega_0)$ having vanished in this region of the $\Delta_1$-$\Delta_2$ phase diagram. Bottom panel: Dispersion relation for the negative imaginary part of $\omega_0$ (left plot) and $\omega_1$ (right plot).}
    \label{fig:excited_3D_plot}
\end{figure}

As soon as we depart from this specific line, we see that a linear component starts developing in the $x$ direction again and the dispersion relation becomes linear in all directions, although increasingly anisotropic as it approaches the red line.  This phase corresponds to the pink region in the phase diagram in figure \ref{fig:phase_diagram}. If we keep increasing $\Delta_2$ we find that $\omega_1$ eventually becomes massless as well 
 and we have a region with two gapless fermions (purple region in the phase diagram in figure \ref{fig:phase_diagram}). Hence in this region we find two modes that disperse linearly as shown in Figure \ref{fig:excited_3D_plot}.
 
 Finally, at the orange line both modes collide and meet, and we find the gapped phase described above.

\label{fig:final_3d_plots}

\section{Conclusions}
\label{sect:conclusion}

In this paper we presented a toy model featuring a topological phase transition between a Dirac and a gapped phase, featuring an anisotropic Dirac semimetal at the critical point. The construction relies on identifying a symmetry breaking pattern on a free field theory with a relativistic UV fixed point. Then, by reproducing such symmetry breaking pattern, we obtained an holographic model consisting on a Schwarchild AdS black hole with a scalar and non abelian fields living outside of it. We test our model by coupling it with probe fermions in the bulk and by computing the fermionic quasinormal modes. The quasinormal frequencies indeed display a qualitative behavior that agrees with the expectations for each of the phases of the model.

As closing remarks, we would like to discuss some open questions. A straightforward extension of our work will be to include backreaction to the model. This would allow us to have a deeper understanding of the dynamics of the system. For instance, we would gain access on the dynamics of the dual stress tensor and in principle we could compute viscosities. Anisotropic Dirac semimetals were reported \cite{link2018out} to violate  the conjectured bound  $\eta/s\geq 1/4\pi$ \cite{Kovtun:2004de} for relativistic holographic systems. It would be interesting to understand if such violation occurs in our holographic model in order to further understand the origin of the tension between these results.
A backreacting model would also allow us to access to lower temperatures and even constructing the corresponding domain walls at $T=0$. In this limit, interesting emerging behaviors were reported previously for topological phase transitions \cite{Landsteiner:2015pdh, Caceres:2023mqz}. These include IR geometries with different Lifshitz scaling in different directions and exotic Boomerang RG flows, where the UV and IR central charges match. Another computation feasible with a backreacting geometry would be to obtain an analog to the pole-skiping phenomena studied in Section \ref{sec:pole_skipping} but now including the effect of the relevant deformations associated to the topological phase transition.

A second possible direction concerns the possibility of having phase transitions. Quantum field theoretical descriptions of Dirac semimetals report a plethora of symmetry breaking patterns at finite interaction strength \cite{Uryszek:2019joy}. It would be interesting to explore if something similar happens in our model that is strongly interacting by construction. In the present construction such phase transitions would correspond to our geometries developing some spontaneous sourceless profiles for some generalized ansatz. AdS black holes with non-Abelian sources are known to be prone to instabilities. The end point of such instabilities is a hairy black hole and many examples appear in the literature \cite{Gubser:2008wv,Juricic:2020sgg,Grandi:2021bsp}. 

A final possible direction we would like to discuss now relies on the possibility of understanding the UV completion of out model. This would allow us to fix the couplings in our theory in a way that is consistent with string theory. In this sense, as our construction mimics the reasoning behind the holographic Weyl semimetal \cite{Landsteiner:2015lsa}, we expect that we might be able to generalize the construction of \cite{BitaghsirFadafan:2020lkh,Matsumoto:2024czp} but now including at least two branes featuring a $2+1$ dimensional intersection.

\section*{Acknowledgments:} This work was supported by ANID/ACT210100 (R.S.-G. and S.B), Fondecyt (Chile) Grant No. 1241033 (R.S.-G.).  I.S.L. thanks ICTP and Universidad Católica de Chile for hospitality during different stages of this project. I.S.L would like to acknowledge support from the ICTP through the Associates Programme (2023-
2028). 

\section*{Appendices}
\appendix
\section{Near-horizon expansion of the Dirac equation and pole-skipping points}\label{sec:appendix_A}
The Dirac equation for the re-scaled, projected spinor fields (eqs.~\eqref{eq:Dirac_eq_1} and \eqref{eq:Dirac_eq_2}) can be analitically expanded around $r = 1$ (the black hole horizon) so as to impose the in-falling boundary conditions, as is presented in eq.~\eqref{eq:half_int_expansion} and restated here for completeness
\begin{equation}\label{eq:half_int_expansion_appendix}
    \zeta_\pm^{\mathrm{IR}}(r) = (1-r)^{-i\omega/4\pi T}\left(a_0^{\pm}+a_{1/2}^{\pm}\sqrt{1-r}+a_{1}(1-r)+a_{3/2}^{\pm}(\sqrt{1-r})^3+a_2^{\pm}(1-r)^2+\cdots\right).
\end{equation}

For the purposes of this section, the background scalar and gauge fields are turned off, which leads to a bulk where only the fermion fields propagate on the gravitational background with no backreaction. In this case, the Dirac equation takes the simpler form
\begin{align}
    \frac{\mathrm{d}\zeta_+}{\mathrm{d}r}+\frac{i}{\sqrt{f(r)}}U \zeta_-&=0 \\
    \frac{\mathrm{d}\zeta_-}{\mathrm{d}r}-\frac{i}{\sqrt{f(r)}}U \zeta_+ & =0 ,
\end{align}
where the matrix operator $U\equiv U(r;\omega,k_x,k_y)$ takes the simpler block-diagonal form
\begin{equation}
    U(r;\omega,k_x,k_y) = \begin{bmatrix}
        k_y & k_x-\frac{\omega}{\sqrt{f(r)}} & 0 & 0 \\
        k_x + \frac{\omega}{\sqrt{f(r)}} & -k_y & 0 & 0\\
        0 & 0 & k_y & k_x-\frac{\omega}{\sqrt{f(r)}} \\
        0 & 0 & k_x+\frac{\omega}{\sqrt{f(r)}} & -k_y
    \end{bmatrix}
\end{equation}
The shape of $U(r;\omega,k_x,k_y)$ implies that the Dirac equations \eqref{eq:Dirac_eq_1},\eqref{eq:Dirac_eq_2} can be simplified from a pair of ODE's that couple two 4-dimensional spinors to a pair of ODE's that couple two 2-dimensional spinors, since the diagonal form of $U$ decouples the lower and upper sectors of $\zeta_\pm$. This is just an indication that the sources and VEV's of the fermion operators on the two graphene layers of the toy-model boundary theory don't sense each other's presence, since the only thing coupling both of them are the gauge and scalar fields. Therefore, we are left with two separate copies of a 2-dimnensional massless Dirac excitation, and therefore the bulk fields also decouple down to two copies of the same ODE's.

Given all of the above, for the purposes of this section the fermion fields $\zeta_\pm(r)$ will be considered to be 2-dimensional spinors, related to each other by the following Dirac equation:
\begin{align}
    \frac{\mathrm{d}\zeta_+}{\mathrm{d}r}+\frac{i}{\sqrt{f(r)}}\begin{bmatrix}
        k_y & k_x-\frac{\omega}{\sqrt{f(r)}}\\
        k_x+\frac{\omega}{\sqrt{f(r)}}&-k_y
    \end{bmatrix}\zeta_- & = 0 \label{eq:Dirac_eq_1_appendix}\\
    \frac{\mathrm{d}\zeta_-}{\mathrm{d}r}-\frac{i}{\sqrt{f(r)}}\begin{bmatrix}
        k_y & k_x-\frac{\omega}{\sqrt{f(r)}}\\
        k_x+\frac{\omega}{\sqrt{f(r)}}&-k_y
    \end{bmatrix}\zeta_+ & = 0 \label{eq:Dirac_eq_2_appendix}
\end{align}
By plugging the expansion \eqref{eq:half_int_expansion_appendix} into \eqref{eq:Dirac_eq_1_appendix} and \eqref{eq:Dirac_eq_2_appendix}, making a series expansion and, finally, solving for each series coefficent order by order, each coefficient $a_j^+$ for $j\in\{1/2,1,3/2,2,\ldots\}$ and $a_j^-$ for $j\in\{0,1/2,1,3/2,2,\ldots\}$ can be solved in terms of $a_0^+$. The set of two numbers contained in $a_0^+ = \begin{bmatrix}
    a_{0,1}^+\\
    a_{0,2}^+
\end{bmatrix}$ correspond to a single initial condition which defines a unique solution to the Dirac equation in the IR, which propagates to the UV through the full numerical solution to the full Dirac equation in the bulk. However, the system of equations that's obtained, order by order, through this procedure involves an explicit parametric dependence on $\omega$, $k_x$ and $k_y$, and therefore there may be special choices of these parameters for which the system of equations (up to certain order) becomes ill-defined \cite{Ceplak:2020}. For these choices of $\omega$, $k_x$ and $k_y$, some coefficients in the expansion \eqref{eq:half_int_expansion_appendix} may not be explicitly expressed as a function of the zeroth-order coefficients $a_0^+$, and a second set of independent initial values for the Dirac equation will arise, making the construction of a unique Green's function on the boundary will not be possible, in principle \cite{Ceplak:2020}.

In order to illustrate the aforementioned procedure for finding the pole-skipping frequencies, let us explicitely state the system of equations for the solution up to order $1/2$ in the Fröbenius expansion \eqref{eq:half_int_expansion_appendix}

\begin{align}
    \omega(a^+_{0,2}-a^-_{0,1})&=0 \label{eq:zeroth_order_eq1_appendix}\\
    \omega(a^+_{0,1}-a^-_{0,2})&=0 \label{eq:zeroth_order_eq2_appendix}\\
    \sqrt{3}(k_ya^+_{0,1}+k_xa^+_{0,2})-\frac{3i}{2}\left(1-\frac{2i\omega}{3}\right)a^-_{1/2,1}-\omega a^+_{1/2,2}&=0 \label{eq:half_order_eq1_appendix}\\
    \sqrt{3}(k_xa^+_{0,1}-k_ya^+_{0,2})-\frac{3i}{2}\left(1-\frac{2i\omega}{3}\right)a^-_{1/2,2}+\omega a^+_{1/2,1}&=0 \label{eq:half_order_eq2_appendix}\\
    \sqrt{3}(k_ya^-_{0,1}+k_xa^-_{0,2})-\frac{3i}{2}\left(1-\frac{2i\omega}{3}\right)a^+_{1/2,1}-\omega a^-_{1/2,2}&=0 \label{eq:half_order_eq3_appendix}\\
    \sqrt{3}(k_xa^-_{0,1}-k_ya^-_{0,2})-\frac{3i}{2}\left(1-\frac{2i\omega}{3}\right)a^+_{1/2,2}+\omega a^-_{1/2,1}&=0 \label{eq:half_order_eq4_appendix}
\end{align}

Solving eqs.~\eqref{eq:zeroth_order_eq1_appendix} and \eqref{eq:zeroth_order_eq2_appendix} will result in $a^-_0$ expresed in terms of $a^+_0$, which (for $\omega\neq 0$) gives the zeroth-order relation $a^-_0 = -i\sigma_2 a^+_0$ (compare to the zeroth-order relation between coeffcients of fermions in the IR presented in \cite{Grandi2022} and \cite{Giordano2017}). This solution is then inserted in eqs.~\eqref{eq:half_order_eq1_appendix} through \eqref{eq:half_order_eq4_appendix}, which would give $a^\pm_{1/2}$ exclusively in terms of $a^+_0$. This last set of equations can be rewritten in matrix form as:
\begin{align}\label{eq:half_order_matrixeq_appendix}
    \begin{bmatrix}
        0 & -\omega & -\frac{3i}{2}\left(1-\frac{2i\omega}{3}\right) & 0 \\
        \omega & 0 & 0 & -\frac{3i}{2}\left(1-\frac{2i\omega}{3}\right) \\
        \frac{3i}{2}\left(1-\frac{2i\omega}{3}\right) & 0 & 0 & -\omega \\
        0 & \frac{3i}{2}\left(1-\frac{2i\omega}{3}\right) & \omega & 0
    \end{bmatrix}\begin{bmatrix}
        a^+_{1/2}\\
        a^-_{1/2}
    \end{bmatrix} = \sqrt{3}\begin{bmatrix}
        k_y & k_x & 0 & 0 \\
        k_x & -k_y & 0 & 0\\
        0 & 0 & k_y & k_x \\
        0 & 0 & k_x & -k_y 
    \end{bmatrix}\begin{bmatrix}
        a^+_0 \\
        a^-_0
    \end{bmatrix},
\end{align}
where $a_0^-$ is implicitly related to $a_0^+$ through $a_0^- = -i\sigma_2 a_0^+$. Equation \eqref{eq:half_order_matrixeq_appendix} has a solution for $a^\pm_{1/2}$ only if the left-hand side matrix is invertible, which fails to be the case when its determinant is zero. Since said determinant is a function of $\omega$, solving for this condition gives that eq.~\eqref{eq:half_order_matrixeq_appendix} will become singular if and only if $\omega = \omega_1 = -\frac{3i}{4}$. This result corresponds precisely to the first fermionic Matsubara frequency outlined in eq.~\eqref{eq:first_pole_skips}.

When $\omega = \omega_1$, the four independent equations \eqref{eq:half_order_eq1_appendix} through \eqref{eq:half_order_eq4_appendix} actually devolve to only two independent equations, which can be taken to be eqs.\eqref{eq:half_order_eq1_appendix} and \eqref{eq:half_order_eq2_appendix}. This system of two equations is solvable, and actually results in a specific linear combination of $a^+_{1/2}$ and $a^-_{1/2}$ that can be expresed in terms of $a^+_0$. When $\omega = \omega_1$, said equations can be rewritten as:
\begin{align}
    a^+_{1/2,1}+a^-_{1/2,2} &= -\frac{4i}{\sqrt{3}}(k_ya^+_{0,1}-k_xa^-_{0,2})\\
    a^+_{1/2,2}-a^-_{1/2,1}&= \frac{4i}{\sqrt{3}}(k_ya^+_{0,1}+k_xa^-_{0,2}),
\end{align}
where the variable $a^-_0$ has been eliminated through the zeroth order relation stated previously. Following \cite{Ceplak:2020}, by calling $b_{1/2} := a^-_{1/2}+i\sigma_2a^+_{1/2}$ one obtains:
\begin{equation}\label{eq:new_half_order_matrixeq_appendix}
    b_{1/2}=\frac{4i}{\sqrt{3}}\begin{bmatrix}
       k_y & k_x \\
       k_x & -k_y
       \end{bmatrix}a^+_0
\end{equation}
Equation \eqref{eq:new_half_order_matrixeq_appendix} states that, even though $a^\pm_{1/2}$ can not each be separately related to the independent initial condition $a^+_0$ of the Dirac equation, a new linear combination of those parameters, called $b_{1/2}$, actually can, and therefore one still has a way of building a unique solution for the IR expansion of the Dirac equation up to order $(1-r)^{1/2}$. However, even this fails if eq.\eqref{eq:new_half_order_matrixeq_appendix} becomes singular, which happens when the right-hand side matrix of this equation has determinant equal to zero. This happens if and only if $k_x = k_y = 0$, which corresponds to the pole-skiping spatial momenta at $\omega = \omega_1$ shown in eq.~\eqref{eq:first_pole_skips}. In this case the $(1-r)^{1/2}$ order contribution to the series expansion \eqref{eq:half_int_expansion_appendix} becomes completely decoupled from the $(1-r)^0$-order part, and therefore both sets of parameters are truly independent. This means that there is complete freedom to choose $b_{1/2}$ independently of $a_0^+$; namely, one can choose $b_{1/2} = 0$ (once again following \cite{Ceplak:2020}). In this case, $a^+_0$ and $a^+_{1/2}$ are both free parameters of the Dirac equation, which set the $a^-_{1/2}$ and $a^-_{0}$ fermion components through:
\begin{equation}
    a^-_0 = -i\sigma_2 a^+_0 \;,\; a^-_{1/2} = -i\sigma_2 a^+_{1/2}
\end{equation}
Given all of the above, there is no unique way of building a solution to the Dirac equation since there are two sets of unrelated free parameters in the near-horizon expansion, and so the boundary Green's function is not uniquely defined \cite{Ceplak:2020}.

Conditions of singularity for higher order coefficients give rise to the rest of fermionic Matsubara frequencies, with their corresponding pole-skipping spatial momenta. As one can see, the pole-skipping condition is entirely determined by the near-horizon behavior of the bulk fields, yet it can be still determined from the near-boundary physics that is associated to the boundary theory through the analytic structure of the Green's function, as it was explicitly shown in Section \ref{sec:pole_skipping}. This is shown back in Figure \ref{fig:pole_skipping_plots}, where the first three Matsubara frequencies, along with their corresponding pole-skipping space momenta, are shown as black dots placed on top of the intersections of lines of poles and zeroes of the Green's function. These black dots indeed correspond to the tuples of frequencies and momenta shown in eq.~\eqref{eq:first_pole_skips}.

\bibliographystyle{JHEP}
\bibliography{references}

\end{document}